\RequirePackage{fix-cm}
\documentclass[smallcondensed]{svjour3}     
\smartqed  
\usepackage{graphicx}
\usepackage{graphicx}
\usepackage{dcolumn}
\usepackage{bm}

\begin{document}           

\title{Particle Pair Production in Cosmological General Relativity}

\author{Firmin J. Oliveira}

\date{Received: date / Accepted: date}

\maketitle

\begin{abstract}

 The Cosmological General Relativity (CGR) of Carmeli, a 5-dimensional (5-D) theory of time,
  space and velocity, predicts the existence of an acceleration $a_0 = c / \tau$ due to the
  expansion of the universe, where $c$ is the speed of light in vacuum, $\tau = 1/h$ is the
  Hubble-Carmeli time constant, where $h$ is the Hubble constant at zero distance and no gravity.
  The Carmeli force on a particle of mass $m$ is $F_c = m a_0$, a fifth force in nature.
  In CGR, the effective mass density  $\rho_{eff} = \rho - \rho_c$, where $\rho$ is the matter
  density and $\rho_c$ is the critical mass density which we identify with the vacuum mass
  density $\rho_{vac} = - \rho_c$.

  The fields resulting from the weak field solution of the Einstein field equations in 5-D CGR 
  and the Carmeli force are used to hypothesize the production of a pair  of particles. 
   The mass of each particle is found to be $m=\tau  c^3 / 4  G$, where $G$ is Newton's  constant.
   The vacuum mass density derived from the physics is $\rho_{vac} =  -\rho_c = -3/8 \pi G \tau^2$. 
   We make a connection between the cosmological constant
   of the Friedmann-Robertson-Walker model and  the vacuum mass density of CGR by the relation
   $\Lambda = -8 \pi G \rho_{vac} = 3 /  \tau^2$.
  Each black hole particle defines its own volume of space enclosed by the event horizon, forming a
  sub-universe.

  The cosmic microwave background (CMB) black body radiation at the temperature
  $T_o=2.72548 \,{\rm K}$ which fills that volume is found to have a relationship
  to the ionization energy of the Hydrogen atom.  Define the radiation energy
  $\epsilon_{\gamma} = ( 1 - g ) m c^2 / N_{\gamma}$,
  where $(1-g)$ is the fraction of the initial energy $m c^2$ which converts to photons,  $g$ is a function of the
   baryon density parameter $\Omega_b$ and $N_{\gamma}$ is the total number
  of photons in the CMB radiation field. We make the connection with the ionization energy of the first
  quantum level of the Hydrogen atom by the hypothesis
  \begin{equation}
    \epsilon_{\gamma} = \frac{ \left( 1 - g \right) m c^2 } { N_{\gamma} } = \frac{\alpha^2 \mu c^2} {2},
         \label{abseq:epsilon_gamma}
  \end{equation}
  where  $\alpha$ is the fine-structure constant and $\mu = m_p  f / (1 + f)$,  where $f= m_e /m_p$  
  with $m_e$ the electron mass and $m_p$ the proton mass. 
  We give a model for $g \approx \Omega_b  ( 1 + f ) m_p  /  m_n $, where $m_n$ is the neutron mass.
  Then ratio $\eta$ of the number of baryons $N_b$ to photons $N_{\gamma}$ is given by
     \begin{equation}
         \eta =  \frac{N_b}  {N_{\gamma}} \approx \frac{ \alpha^2   \Omega_b  f m_p / m_n} { 2 \left( 1 + f \right) 
            \left[ 1 - \Omega_b \left( 1 + f  \right) m_p  / m_n \right] }
              \label{abseq:N_n-over_N_gamma}
      \end{equation}
  with a value of $\eta  \approx 6.708 \times 10^{-10}$.

 The Bekenstein-Hawking black hole entropy $S$ is given by $ S = ( k c^3 A ) / ( 4 \hbar G )$, where $k$ is Boltzmann's constant,
 $\hbar$ is Planck's constant over $2 \pi$ and  $A$ is the area of the event horizon.
  For our black hole  sub-universe  of mass $m$ the entropy is given by
  \begin{equation}
           S = \frac{\pi  k \tau^2 c^5 } { \hbar G },
 \end{equation}
 which can be put into the form relating to the vacuum mass density
 \begin{equation}
      \rho_{vac} =  \frac{\rho_P } { \left( S / k \right) },
 \end{equation}
  where  the cosmological Planck mass density $\rho_P = -{\cal M}_P /  L^3_ P$. The  cosmological  Planck mass
   ${\cal M}_P = \sqrt{ \sqrt{ 3 / 8 } \, \hbar c / G}$  and  length $L_P = \hbar /   {\cal M}_P c$.
   The value of $ (S/k) \approx 1.980 \times 10^{122}$.

\end{abstract}

\section{Introduction}
   Carmeli's Cosmological General Relativity(CGR), \cite{behar-carmeli} and \cite{carmeli-1,carmeli-1.1},  is a
   5-dimensional (5-D) theory of  time, space and velocity. Carmeli predicted \cite{carmeli-2} an
   acceleration due to the expansion of the universe
   \begin{equation}
     a_0 = \frac{c}{\tau}, \label{eq:a_0}
   \end{equation} 
   where $c$ is the speed of light in vacuum and $\tau = 1/h$ is the Hubble-Carmeli time
   constant where $h$ is the Hubble constant at zero distance and no gravity.
   The Carmeli force on a particle of mass $m$ is $F_c = m a_0$, a fifth force in nature.
   As of this writing the published values \cite{hartnett-0} for 
   $\tau = 4.28 \times 10^{17} {\rm s} = 13.58 \;{\rm Gyr}$
   and $h = 1/\tau = 72.17 \pm 0.84 \;{\rm km} \;{\rm s}^{-1} \;{\rm Mpc}^{-1}$.
   In this paper, unless noted otherwise, all calculations have used this published value for $\tau$.

    We use the weak field solution of CGR to derive $a_0$. 
   The fields and the Carmeli force are used to hypothesize the production of a particle and its
   antiparticle.   The vacuum mass density  $\rho_{vac}$ is defined.
   By applying the physical laws,  the masses and separation distance are determined for the particles.
   Consequently, a value for $\rho_c = - \rho_{vac}$ is obtained, which is derived from the physics  rather than
   being assumed as was previously done.

 \section{Geodesic equations in 5-D}

  The coordinates in 5-D CGR are $t$ for time, $(x^1, x^2, x^3)$ for space, and $v$ for
  expansion velocity.  The line element for a linearized metric is given by
  \begin{eqnarray}
   ds^2 &=& (1 + \phi(r)/c^2) (dx^0)^2  \nonumber \\
             & &  -  \, (dx^1)^2 - (dx^2)^2 - (dx^3)^2   \nonumber \\
             & &  + \, (1 + \psi(r)/\tau^2) (dx^4)^2  , \label{eq:ds2_5D}
  \end{eqnarray}
   where $dx^0 = c dt$, $c$ is the speed of light in vacuo, $dx^4 = \tau dv$ and
   $\tau$ is the Hubble-Carmeli time constant. The spatial line element
  \begin{equation} 
  r = \sqrt{ ( x^1 )^2 + ( x^2 )^2 + ( x^3 )^2 }.   \label{eq:line_element_r}
  \end{equation}

  The geodesic equation is expressed
  \begin{eqnarray}
    \frac{d^2 x^{\mu}}{ds^2} + \Gamma^{\mu}_{\alpha \beta} \frac{dx^{\alpha}}{ds} \frac{dx^{\beta}}{ds} &=& 0
           , \label{eq:geodesic} \\
  \nonumber \\
    \alpha, \beta, \mu &=& 0, 1, 2, 3, 4  ,
  \end{eqnarray}
  where $\Gamma^{\mu}_{\alpha \beta}$ are the Christoffel symbols and repeated indices are summed
  (Einstein summation convention.)  This describes five equations of motion.
  By changing the independent parameter $s$ into an arbitray parameter $\sigma$, and  then letting $\sigma$ 
  be the coordinate $x^1$ in one instance and  the coordinate $x^4$
  in another instance, as detailed in \cite{carmeli-2}, and then using the Einstein-Infeld-Hoffmann method
  in the first approximation,  (\ref{eq:geodesic}) reduces to
  \begin{eqnarray}
   \frac{d^2 x^k}{dt^2} &=& - \frac{1}{2}\frac{\partial{\phi(r)}}{\partial{x^k}}  , \label{eq:d2x/dt2} \\
 \nonumber \\
   \frac{d^2 v}{dt^2} &=& 0  , \label{eq:d2v/dt2} \\
 \nonumber \\
   \frac{d^2 x^k}{dv^2} &=& - \frac{1}{2}\frac{\partial{\psi(r)}}{\partial{x^k}}  , \label{eq:d2x/dv2} \\
 \nonumber \\
   \frac{d^2 t}{dv^2} &=& 0  , \label{eq:d2t/dv2} \\
 \nonumber \\
    k &=& 1, 2, 3  .
  \end{eqnarray}
 
   The first integral of (\ref{eq:d2v/dt2}) gives
  \begin{equation}
    \frac{dv}{dt} = a_0  , \label{eq:dv/dt=a0}
  \end{equation}
  where $a_0$ is a constant.
  The first integral of (\ref{eq:d2t/dv2}), assuming that $a_0 \ne 0$,
  gives
  \begin{equation}
    \frac{dt}{dv} = \frac{1}{dv/dt} = a^{-1}_{0}.  \label{eq:dt/dv=1/a0}
 \end{equation}
  Then (\ref{eq:d2x/dv2}) can be expressed
  \begin{eqnarray}
    \frac{d^2 x^k}{dv^2} &=& \left[ \frac{d}{dt} \frac{dx^k}{dv} \right]
               \frac{dt}{dv}, \\
 \nonumber \\
          &=& \left( \frac{d^2 x^k}{dt^2} \right) \left( \frac{dt}{dv} \right)^{2} , 
                 \label{eq: d2x/dv2_D} \\
  \nonumber \\
         &=& \left( \frac{1}{a^2_0 } \right) \left( \frac{d^2 x^k}{dt^2} \right),
          \label{eq:d2x/dv2=a0^{-2}*d2x/dt2}
  \end{eqnarray}
  where we have substituted from (\ref{eq:dt/dv=1/a0}) for $dt/dv$ in (\ref{eq: d2x/dv2_D}) to get
  (\ref{eq:d2x/dv2=a0^{-2}*d2x/dt2}).
  Finally, from (\ref{eq:d2x/dt2}), (\ref{eq:d2x/dv2}) and (\ref{eq:d2x/dv2=a0^{-2}*d2x/dt2}) we get 
  \begin{equation}
    \frac{\partial{\psi(r)}}{\partial{x^k}} =  \left( \frac{1}{a^2_0 } \right)
           \frac{\partial{\phi(r)}}{\partial{x^k}}.  \label{eq:partial(phi)=partial(psi)}
  \end{equation}
  To show that $\psi(r) = (\tau^2/c^2)\phi(r)$, which from (\ref{eq:partial(phi)=partial(psi)}) implies that
  $a_0 = \pm c / \tau$,  the solutions to the field equations are needed.  

 \section{Einstein field equations in 5-D}

  The Einstein field equations in 5-D are
  \begin{eqnarray}
    R_{\mu \nu} - \frac{1}{2} g_{\mu \nu} R &=& \kappa  T_{\mu \nu},   \label{eq:EinsteinEqn5D} \\
 \nonumber \\
    \mu, \nu &=& 0, 1, 2, 3, 4,
  \end{eqnarray}
  where $R_{\mu \nu}$ is the Ricci tensor, $R$ its contraction, $g_{\mu \nu}$ is the metric,
  $\kappa=8 \pi G / c^2 \tau^2$ and the stress-energy tensor is given by
  \begin{equation}
     T_{\mu \nu} = \rho_{eff}  u_{\mu} u_{\nu} + p \left(u_{\mu} u_{\nu} - g_{\mu \nu} \right),
       \label{eq:Tstress}
  \end{equation}
  where
  \begin{equation}
    \rho_{eff} = \rho - \rho_c  \label{eq:rho_eff}
  \end{equation}
  is the effective mass density, with $\rho$ the matter density and $\rho_c$ the critical mass density.
  Also, $p$ is the pressure and $u_{\mu}=u^{\mu} = (1,0,0,0,1)$ is the velocity vector.
  Solving \cite{carmeli-2,carmeli-1,carmeli-1.1} the field equations (\ref{eq:EinsteinEqn5D}) for the metric defined by
 (\ref{eq:ds2_5D}), the equations for the fields $\phi(r)$ and $\psi(r)$ are found to be
  \begin{equation}
    \nabla^2{\phi(r)} = \left( c^2/\tau^2 \right) \nabla^2{\psi(r)} = -16 \pi G \rho_{eff}  , \label{eq:nabla2_phi_psi}
  \end{equation}
  The solutions to (\ref{eq:nabla2_phi_psi}) for the non-homogeneous case are given by
  \begin{equation}
    \phi(r) = \left( c^2 / \tau^2 \right) \psi(r)
            = - \rho_{eff} \left( \frac{8 \pi G}{3} \right) r^2,
                \label{eq:phi_1_and_psi_1_nonhomogen}
  \end{equation}
  and for the homogeneous case by
  \begin{equation}
    \phi(r) = \left( c^2 / \tau^2 \right) \psi(r) = -\frac{2 G m}{r}, \label{eq:phi_1_and_psi_1_homogen}
  \end{equation}
  where $m$ is a constant mass.
  The general solution is the sum of (\ref{eq:phi_1_and_psi_1_nonhomogen}) and (\ref{eq:phi_1_and_psi_1_homogen}),
  \begin{eqnarray}
   \phi(r) &=& -\rho_{eff} \left( \frac{8 \pi G}{3} \right) r^2  - \frac{2 G m}{r},  \label{eq:phi_general_sol} \\
   \nonumber \\
   \psi(r) &=& -\rho_{eff} \left( \frac{8 \pi G}{3} \right) \frac{\tau^2  r^2} {c^2}  - \frac{2 G m \tau^2}{c^2 r}.  \label{eq:psi_general_sol}
 \end{eqnarray}

  \section{The Carmeli force}

  By combining the three spatial coordinates $(x^1, x^2, x^2)$ into radial  $r$ in
  (\ref{eq:partial(phi)=partial(psi)})
  and by (\ref{eq:phi_1_and_psi_1_nonhomogen}) and (\ref{eq:phi_1_and_psi_1_homogen}),
  \begin{equation}
   \nabla{\psi(r)} =  \left( \frac{\tau^2}{c^2} \right) \nabla{\phi(r)}.
            \label{eq:nabla(psi)=a0^2 nable(phi)}
  \end{equation}
  Assuming $\nabla{\psi(r)} \ne 0$, (\ref{eq:dv/dt=a0}), (\ref{eq:partial(phi)=partial(psi)}) and
  (\ref{eq:nabla(psi)=a0^2 nable(phi)}) imply
  \begin{eqnarray}
     a^2_0 = c^2 / \tau^2  , \label{eq:a0^2=c^2/tau^2} \\
 \nonumber \\
     \frac{dv}{dt} = a_0 = \pm c / \tau  . \label{eq:a0=c/tau}
  \end{eqnarray}

  In summary, the 5-D geometrical system implies the geodesic equations
  (\ref{eq:d2x/dt2})-(\ref{eq:d2t/dv2}).  They indicate that,  for very great distances
  from a source of mass, where the acceleration $\nabla{\phi(r)} \approx 0$ though not zero, there
  is still the effect of the constant acceleration $a_0$ caused by expansion in the velocity dimension
  which is directed along the radial spatial direction. The Carmeli force on a particle of mass $m$
  in vector notation is given by
  \begin{equation}
    {\bf F_c} = m a_0 {\bf r} / r.  \label{eq:F_carmeli}
  \end{equation}
  Hence, the resulting time rate of change in displacement ${\bf r}$ of a particle in 5-D is
  expressed\cite{carmeli-2} in vector notation by
  \begin{equation}
   \frac{d^2 {\bf r}}{dt^2} = -\nabla{\phi(r)} + a_0 {\bf r} / r  ,
          \label{eq:d^2 r/dt^2 =-nabla(phi(r))}
  \end{equation}
  where the actual value of $a_0$, either $+c/\tau$ or $-c/\tau$ is determined by the physics.
  Note that the Carmeli force occurs only when the gradient of the fields $\phi(r) $ and $\psi(r)$ are non-zero.
  Since the fields propagate at speed $c$, so must the Carmeli force field.

  \section{Particle pair production}

   In 5-D CGR there are only two parameter constants, a maximum speed $c$, the speed
   of light in vacuum, and a maximum time $\tau$, the age of the universe.  In the following
   discourse, we use the term {\em matter} in a general way to include all particles of non-zero and
   zero rest mass.  Likewise, we use the term {\em vacuum} to refer to that which is unmaterialized
   but which has energy.

  By the definition of $\rho_{eff} = \rho - \rho_c$  found in  \cite{behar-carmeli} and  \cite{carmeli-1}, Sect. 4.3.1.,
  we have the difference of two positive quantities $\rho$ and $\rho_c$.  It is common in CGR to 
  associate $\rho_c$ with $\rho_{vac}$\cite{carmeli-kuzmenko,hartnett-0}.  If the vacuum density equals
  the critical density, $\rho_{vac} = \rho_c$, as is the usual definition,  then the effective mass density would be
  $\rho_{eff} = \rho - \rho_{vac}$.
  But what is the physical process that requires one to  subtract the densities? Is it not more fundamental to 
  define the effective mass density $\rho_{eff}$ as the sum of all types of mass densities positive and negative?
  In this way we can look at the vacuum density as a negative mass density which is added to the positive
  mass density to obtain an effective density.  Even though the original intent may not have been just this
  idea, we are lead to this point of view by the physical concept. We will show subsequently that the vacuum 
  can indeed be described by a negative mass density when we consider entropy.   Hence, in this paper,
  the effective mass density $ \rho_{eff}$  is defined by
    \begin{equation}
     \rho_{eff} =  \rho + \rho_{vac},  \label{eq:rho_eff=rho+rho_vac}
   \end{equation}
  where $\rho_c = -\rho_{vac}$.

   Before any particles and vacuum were produced the universe is assumed to have been empty with zero
   total energy. Thus, by energy conservation the total energy must always be zero. Suppose that a pair of
   electrically neutral particles is produced along with the vacuum.  By the laws of quantum physics each particle
   must be the antiparticle of the other one in order to conserve all quantum numbers.  This means, among 
   other things, that each particle has the same mass m.  For clarity we assume that one of the particles is
   composed of $(x + y)  q$ amount of matter and $x{\bar q}$ amount of antimatter such that 
   $m = x(q+{\bar q}) + y q$  and the antiparticle ${\bar m} = x(q+{\bar q}) + y {\bar q}$.
   Summing the masses produces $m + {\bar m} = (2x + y) q + (2x + y){\bar q}$, showing an equal
   amount of matter and antimatter for the two particles combined and all quantum numbers are conserved.
 
   Assume that the particles are produced with zero relative velocity ( no kinetic energy) at a distance $r_0$ apart,
   with the vacuum filling the finite universe.   We make our observations from a reference frame with origin
   fixed to the center of one of the particles.  Initially, at the time in the past  $t=\tau$ when the particles
   and vacuum are produced, before the gravitational fields of the particles have begun to propagate at the 
   speed $c$, the only energies are that of the two  rest  masses and the vacuum, so these must sum to zero 
   by  energy conservation.\footnote{ If the initial energy was non-zero, the non-zero amount  would have to be in the 
       form of mass or vacuum since these are the only physical forms we know about. Suppose the initial matter mass and 
       vacuum mass were given by $2 m + m_{vac} = m_i  + m_{vac_i}$.  Since the assumed model  is only for a
       two particle and vacuum system, we could bring everything over to the left hand side and have the new initial condition be
       $2(m  -  m_i/2)  + (m_{vac} - m_{vac_i})  = 0$.  Hence, the initial energy being zero is a logical result of the assumed model.}  
   This is the energy  from a {\em global}  perspective.  The particles occupy the lowest energy state of the system.
   It is assumed that the particles  and the vacuum are produced 
   instantaneously. Hence,  the total energy of the two particles and the vacuum is
   \begin{equation}
     2 m  c^2 +  \rho_{vac} \left(\frac{4}{3} \pi R^3_{S} \right) c^2 = 0 ,
         \label{eq:mass_energy = vacuum_energy}
   \end{equation} 
   where $ R_{S}$ is the (Schwarzschild) radius of the universe and we used the flat space Euclidean
   volume.  Given this radius, the two particle matter density is given by
   \begin{equation}
     \rho = \frac{ 2  m}{(4/3) \pi R^3_{S}}.  \label{eq:rho_define}
   \end{equation}
   Eqs. (\ref{eq:rho_eff=rho+rho_vac})-(\ref{eq:rho_define}) imply that initially
   \begin{equation}
       \rho_{eff} =  \rho + \rho_{vac} = 0.  \label{eq:rho_eff_initial}
   \end{equation}
   In CGR, (\ref{eq:rho_eff_initial})  is the condition for flat space, which implies that Newton's laws are
   valid as a first  approximation.   Also, since $\rho_{eff} = 0$,  there is only the homogeneous solution
   to consider  for (\ref{eq:nabla2_phi_psi}) which is given by (\ref{eq:phi_1_and_psi_1_homogen}).

   The next calculation is made when the field from the remote particle has arrived at the origin
   of the local particle reference system. 
   Although the gravitational field and the Carmeli force field will take a long time to propagate at 
   speed $c$ over distance $r_0$, 
   we do our calculations just when it has arrived at time $t = \tau - r_0 / c$.   (To make this evaluation
   after billions of years have elapsed we must assume, just for this calculation, that the  particles
   are absolutely stable.) 
   Prior to the field arrival there is  no net force acting on either particle.
   Analysis is made relative to the line $r$ joining the particles
   and it is assumed that the initial forces on each particle will be small enough so that we can make the
   weak field approximation for 5-D CGR.

   We analyze the forces upon and energy of the particle in our local system. The forces are due to the
   Newtonian potential and the Carmeli force. The energies are potential energy, kinetic energy (assumed zero), and
   rest energy.  Define the effective potential energy function for the particle comprised of
   gravitation and Carmeli terms,
   \begin{equation}
    \Phi_{eff}(r) = -m^2 G / r  - m  a_0  r. \label{eq:Phi_eff(r)}
   \end{equation}

   Taking the negative gradient of this potential, the total force ${\bf F}(r)$ on the particle is
   \begin{equation}
     {\bf F}(r) = -{\bf \nabla}{\Phi_{eff}(r)} = -\frac{m^2 G}{r^2} \hat{\bf r} + m  a_0 \hat{\bf r},
             \label{eq:F(r)}
   \end{equation}
   where $\hat{\bf r} = {\bf r}/r$, ${\bf r} = (x^1, x^2, x^3)$.
   By Newton's third law, since there are no other forces, the forces due to gravity and the 
   Carmeli force cancel at $r = r_0$,
   \begin{equation}
      {\bf F}(r_0) = 0 \hat{\bf r}. \label{eq:F=0}
   \end{equation}

   The total energy $E(r)$ for the particle at rest is
   \begin{equation}
     E(r) = \Phi_{eff}(r) + m  c^2 =  -m^2 G / r  - m  a_0  r  + m  c^2,
             \label{eq:E(r)}
   \end{equation}
   where we have added the rest energy as the integral constant.   The energy is zero at location $r_0$,
   \begin{equation}
      E(r_0) = 0.  \label{eq:E=0}
   \end{equation}

   With $a_0 = + c / \tau$, (\ref{eq:F(r)})-(\ref{eq:E=0}) can be solved for $m$ and $r_0$
   yielding
   \begin{eqnarray}
     m &=& \frac{\tau  c^3}{4  G}  ,  \label{eq:m_solution} \\
  \nonumber \\
     r_0 &=& \frac{c  \tau}{2}   .  \label{eq:r_solution}
   \end{eqnarray}
 
   Considering the volume containing both particles of mass $2 m$, the  Schwarzschild  radius $R_{S}$ is
   \begin{equation}
     R_{S} = \frac{2 G \left(2 m \right) }{c^2} = \frac{2 G}{c^2} \left(\frac{ \tau  c^3}{2  G} \right) = c \tau.
            \label{eq:R_2S}
   \end{equation}
   The volume $V_S$ of the combined masses is
    \begin{equation}
       V_S = \frac{4 \pi} {3} c^3 \tau^3.  \label{eq:Volume_S}
    \end{equation}
   In section  (\ref{sec:particle-energy}) we will determine the volume of the hemisphere of one of the black holes
   and it will be seen to be half of the volume used in (\ref{eq:Volume_S}).
   With the radius (\ref{eq:R_2S}) in  (\ref{eq:rho_define}) we get the (two particle)  matter density
   \begin{equation}
     \rho =  \frac{3}{8 \pi G \tau^2},
      \label{eq:rho_value}
   \end{equation}
   and by (\ref{eq:rho_eff_initial}) the vacuum mass density
   \begin{equation}
      \rho_{vac} = \frac{-3}{8 \pi G \tau^2}.  \label{eq:rho_vac_value} 
   \end{equation}
  With $\tau = 4.28 \times 10^{17} {\rm s}$ and the standard value for  $G$,  the value
   of $\rho_c = -\rho_{vac} = 3 / 8 \pi G \tau^2 \approx 9.77 \times  10^{-30} {\rm gm / cm^3}$
   is obtained from the physics rather than by assumption. The particle mass
    $ m \approx  4.32 \times 10^{55} {\rm gm}  \approx 2.17 \times 10^{22} \, {\rm M_{\odot}} $.
   Since the particles are each so massive we conclude that each is a black hole  \cite{costa-1}.
   Furthermore,  by the laws of particle physics, each must be the antiparticle of the other.

   There is no cosmological constant in CGR. However, we can relate the CGR vacuum density to the
   cosmological constant $\Lambda$ of standard Friedmann-Robertson-Walker (FRW) cosmology
   \cite{carmeli-kuzmenko}  by
   \begin{equation}
     \Lambda = -8 \pi G \rho_{vac} = \frac{3}{\tau^2},  \label{eq:cosmo_constant}
   \end{equation}
    where the minus sign is required because in this paper the vacuum density is negative.  From this
    we get a value of $\Lambda \approx 1.64 \times 10^{-35} s^{-2}$.

\section{Energies, particle counts and constants  } \label{sec:particle-energy}

   Let us look at energies and particle numbers.
   The mean total energy density in all frequencies of the radiation in a black body at a  temperature $T$ is 
    well known\cite{reif-1} and given by
    \begin{equation}
       u(T) = \frac{\pi^2 k^4 T^4} {15 c^3 \hbar^3}.  \label{eq:u(T)}
    \end{equation}

    Consider the volume $V_m$ of the sub-universe of one of the black hole particles of mass $m$ given by (\ref{eq:m_solution}).
    Since the interacting black hole event horizons are so close together they are deformed as described by Fig. 2 (c) of
    \cite{costa-1}.  Referring to  Fig. 1 of \cite{costa-1},  $z_1 - z_2 = c \tau / 2$, $(\mu_1 + \mu_2)/2 = c \tau /2$,
    giving $\Delta{z} =  z_1 - z_2  -  (\mu_1 + \mu_2)/2 = 0$. For this extreme case the black hole volume
    is a hemisphere with radius
    \begin{equation}
      R_m =  \frac{ \left( 2 \right) 2 G m } {  c^2 }  = c  \tau.   \label{eq:Rm}
    \end{equation}
    Then the hemispheric volume of the sub-universe is $V_m = (2/3) \pi R^3_m = (2/3) \pi c^3 \tau^3$.
    Then the total energy $E_{cmb}$  of the cosmic microwave background (CMB) radiation (\ref{eq:u(T)})
    at the  present CMB temperature $T_o$  is given by
    \begin{equation}
       E_{cmb} = u(T_o) \, V_m = \frac{2 \pi^3 \tau^3 k^4  T^4_o} {45 \hbar^3}.  \label{eq:E_cmb}
    \end{equation}
    For the current CMB temperature $T_o=2.72548 \,{\rm K}$  we get a total energy of
   $E_{cmb} = 1.85 \times 10^{72} {\rm erg}$  which is equivalent to  $1.03 \times 10^{18} {\rm M_{\odot}}$.

    The mean energy of the photons in a black body at temperature $T$ is given by $\zeta \,k \,T$,
     where $\zeta = 2.70117803$ is the black body mean energy coefficient\cite{akerlof-1}. 
    The number $N_{\gamma}$ of photons in the CMB at the current temperature $T_o$ is then given by
    \begin{equation}
       N_{\gamma} = \frac{E_{cmb}}{\zeta \,k \,T_o} \approx 1.82 \times 10^{87}.   \label{eq:N_gamma}
    \end{equation}

 \subsection{Black body at the ionization epoch} \label{sec:black_body_ionize}

    An interesting discovery  is made by asking what was the average energy $\epsilon_{\gamma}$ of the CMB photons when
    they had a total energy which was equal to the initial particle rest energy $m c^2$?  Specifically, that  average energy
    is given by
    \begin{equation}
       \epsilon_{\gamma} \approx \frac{m c^2} {N_{\gamma}} \approx 13.32 \, {\rm eV},  \label{eq:E_gamma}
     \end{equation}
    which is $98 \,\%$ of the ionization energy $13.6 \,{\rm eV}$ of the Hydrogen atom.
    Since the CMB is thought to be directly related to
    the ionization of the Hydrogen atom, it is plausible that $ \epsilon_{\gamma}$, when defined properly,
    is exactly equal to the ionization energy  of the first quantum level of the Hydrogen atom.
    We make the hypothesis, with justification given in section (\ref{sec:ratio_N_n_to_N_gamma}),  that
    \begin{equation}
        \epsilon_{\gamma} =  \frac{\left( 1 - g \right) m c^2} {N_{\gamma}} = \frac{\alpha^2 \mu c^2} {2},  \label{eq:epsilon_gamma_H_ionize}
    \end{equation}
   where $(1-g)$ is the fraction of the initial energy $m c^2$ which converts to photons during the big bang,  $g$ is a function of the
   baryon density parameter $\Omega_b$,
  $\alpha$ is the fine-structure constant and $\mu = m_e  m_p / (m_e + m_p)$ is the reduced electron mass    
   in the Hydrogen atom, with $m_e$ and $m_p$ the masses of the electron and proton, respectively.
   Combining (\ref{eq:m_solution}), (\ref{eq:E_cmb}),  (\ref{eq:N_gamma}) and (\ref{eq:epsilon_gamma_H_ionize})
  and simplifying we obtain
   \begin{equation}
       \epsilon_{\gamma} = \frac{45 \zeta \left( 1 - g \right) c^5 \hbar^3} {8 \pi^3 G \tau^2  k^3  T^3_o} 
                = \frac{\alpha^2 \mu c^2} {2}.  \label{eq:epsilon_gamma_H_ionize_expanded}
   \end{equation}
   When the temperature of the universe was at the  $13.6 \, {\rm eV}$ energy level it is expected that by and large there were
   no more baryon or lepton productions.   We therefore model the plasma as containing only protons, electrons, neutrinos and 
   photons in interaction.
    In this plasma phase the average kinetic energy is small compared to the rest energies of the proton and electron,
    so we will ignore the neutrino in this model, as a first approximation.   Then we can say that
    \begin{equation}
      N_{\gamma} \epsilon_{\gamma} \approx m c^2 - g_b N \left( m_p + m_e \right) c^2,  \label{eq:Ngamma_Egamma}
    \end{equation}
    where  we assume that $N = m / m_n$ is the number of baryons in the original mass $m$ with $m_n$ the neutron mass,
    $g_b= N_b / N$ is the fraction of baryons (and leptons) which survived the annihilations during the big bang and
    $N_b$ is the number of baryons after the annihilations. Substituting these values for $m$ and $N$
    in (\ref{eq:Ngamma_Egamma}) and simplifying we have
    \begin{equation}
        N_{\gamma} \epsilon_{\gamma} \approx m c^2  \left[1 - g_b \left( \frac{m_p + m_e } { m_n} \right) \right].
            \label{eq:Ngamma_Egamma_simp}
    \end{equation}
    The remnant baryonic mass fraction $g_b$  has the relationship to (\ref{eq:rho_value}),  the original matter density $\rho$, by
     \begin{equation}
       g_b  = \frac{N_b m_n} { m }  = \frac{N_b m_n } {\rho  V_m}  = \left( \frac{N_b m_n} { V_m} \right)  \frac{1} { \rho}
            = \frac{\rho_b} { \rho },   \label{eq:g_rho_b}
     \end{equation}
    where $\rho_b$ is the baryon matter density at the current epoch, $\rho$ is the original matter density (\ref{eq:rho_value})
    and $V_m$ is the sub-universe volume.  Since the original matter density equals the critical density  $\rho = \rho_c$, then 
     \begin{equation}
      g_b = \frac{ \rho_b } { \rho_c } = \Omega_b,  \label{eq:g_Omega_b}
     \end{equation}     
     where $\Omega_b$ is the baryon matter density parameter at the current epoch. 
    Comparing (\ref{eq:Ngamma_Egamma_simp})  and (\ref{eq:epsilon_gamma_H_ionize}), and substituting
    for $g_b$ from  (\ref{eq:g_Omega_b})  we have
    \begin{equation}
      g \approx \Omega_b \left( \frac{ m_p + m_e } { m_n } \right).   \label{eq:g_gb}
    \end{equation}
    Substituting values for the parameters in (\ref{eq:g_gb}),
     with $\Omega_b = 0.0443$ from \cite{lambda-1} we have
     \begin{equation}
        g =  0.04426.   \label{eq:g_value}
     \end{equation}
  The fact that $g_b = \Omega_b$ is consitent with the findings of \cite{hartnett-0} where it was reported that 
  the matter density $\Omega_m \approx \Omega_b$.

  For the radiation field at the time of the Hydrogen ionization epoch of the big bang, assuming that it
  was a black body, the average energy would be $\epsilon_{\gamma}  = \zeta \, k \, T_{\gamma} = \alpha^2 \mu c^2 /  2$,
  from which we get the ionization black body temperature
  \begin{equation}
     T_{\gamma} = \frac{ \alpha^2 \mu \,c^2} { 2 \zeta \,k} \approx  5.84 \times 10^4 {\rm K}.  \label{eq:T_gamma}
  \end{equation}  
   The Hubble-Carmeli time constant $\tau$ is difficult to measure so let us find its value in terms of all
   the other parameters which, except for $g$, we know more precisely.  Inverting (\ref{eq:epsilon_gamma_H_ionize_expanded})
   to obtain $\tau$ we have
   \begin{equation}
      \tau = \sqrt{ \frac{ 45 \zeta \left( 1 - g \right) c^3 \hbar^3} {4 \pi^3 G \mu \alpha^2 k^3 T^3_o}    }.  \label{eq:tau_formula}
   \end{equation}
   Substituting values for the parameters in (\ref{eq:tau_formula}), with $g = 0.04426$ at the current epoch,
   we obtain $\tau \approx 4.144 \times 10^{17} {\rm s}$ which corresponds to $h = 74.47 \, {\rm km} \, {\rm s}^{-1} {\rm Mpc}^{-1}$.

    Given that, the cosmological constant (\ref{eq:cosmo_constant})
    \begin{equation}
        \Lambda = \frac{3}{\tau^2} = \frac{4 \pi^3 G \mu \alpha^2  k^3 T^3_o}  {15 \zeta \left( 1 - g \right) c^3 \hbar^3}.
              \label{eq:lambda_formula}
    \end{equation}
    With this expression and again with $g = 0.04426$, $ \Lambda \approx 1.747 \times 10^{-35} s^{-2}$.

   The initial volume of the black hole sub-universe is $V_m$.
    It is possible that the black holes sub-universes are in a static condition as described in \cite{costa-1}.
    The fact that we get good results with very little tweaking of the single parameter $\tau$ is amazing to say the least.
   Tentatively, this could suggest that the expansion of the universe is actually the filling of the static volume $V_m$ of the sub-universe 
    with radiation and matter from the Hydrogen ionization era as well as from all subsequent radiations from galaxies, stars, etc. 
    In cosmological special relativity (\cite{carmeli-1}, p. 19), the cosmic time is a relative concept. Radiation of wavelength
   $\lambda_e$  emitted at a cosmic time $t$ in the past will be redshifted to wavelength $\lambda_o$ at the present epoch time $0$
   by the relation
    \begin{equation}
       \frac{\lambda_o} {  \lambda_e } = \sqrt{ \frac{ 1 + t/\tau} { 1 - t/\tau} }.     \label{eq:cosmological_redshift}
    \end{equation}

\subsection{Ratio of the number of baryons to  CMB photons} \label{sec:ratio_N_n_to_N_gamma}

    We want to substantiate the claim made in (\ref{eq:epsilon_gamma_H_ionize}).
    Using (\ref{eq:Ngamma_Egamma}), suppose we re-write it in the form 
    \begin{equation}
        \epsilon_{\gamma} \approx  \left( \frac{ N_b m_n c^2 }  {g_b  N_{\gamma}}  \right) \left[1 - g_b \left( \frac{m_p + m_e } { m_n} \right) \right] 
              = w \frac{\alpha^2 \mu c^2} {2},
              \label{eq:w_epsilon_gamma_H_ratio}
    \end{equation}
    where we substituted for $m = N_b m_n / g_b$,
    and the factor $w$  allows us to assert that the statement is certainly true for some $w$.
    Assume $w=1$. 

     Express the electron mass as a fraction $f$ of the proton mass by  $m_e = f m_p$. 
     Then the  reduced mass of the electron in the Hydrogen atom is
     \begin{equation}
        \mu = \frac{f m_p}{ \left( 1 + f \right)}.  \label{eq:mu_fmp}
     \end{equation}
    Solving (\ref{eq:w_epsilon_gamma_H_ratio}) for $\eta = N_b / N_{\gamma}$ and substituting  $\Omega_b = g_b$ from (\ref{eq:g_Omega_b})
    we have
      \begin{equation}
         \eta =  \frac{N_b}  {N_{\gamma}} \approx \frac{ \alpha^2   \Omega_b  f m_p / m_n} { 2 \left( 1 + f \right) 
           \left[ 1 - \Omega_b \left( 1 + f  \right) m_p  / m_n \right] }.
              \label{eq:N_n-over_N_gamma}
      \end{equation}
     Substituting values in (\ref{eq:N_n-over_N_gamma}), with $\Omega_b = 0.0443$ from \cite{lambda-1},
     gives the ratio of baryons to photons 
     $\eta  \approx 6.708 \times 10^{-10}$, which is just outside of the range of values ($\eta_{10} = 4.7 - 6.5$)
     from the big-bang nucleosynthesis review\cite{fields}.

     From the previous section (\ref{sec:black_body_ionize})  the ratio of ionization epoch model prediction to
     theoretical value was $0.98 = 1 - (13.6-13.32 ) / 13.6$.
     For the baryon to photon number ratio the model prediction to big bang nucleosynthesis predition 
     was $0.97 = 1 - (6.708 - 6.5)/ 6.5$.  The combined confidence is  $ 0.95 = 0.98 \times 0.97$ that 
     the value of $w = 1$, roughly speaking.

\subsection{Size of the black body at the ionization epoch}

   We want to determine the physical volume $V_{\gamma}$ of the black body which enclosed
   the cosmic background radiation (CBR) at the ionization epoch when it was at the higher temperature $T_{\gamma}$.
   We assume the following basic model for the transformation of the energy from one temperature epoch to another,
    \begin{equation}
        \frac{E_i} {V_i}  = f(T_i, T_f) \frac{E_f} {V_f},  \label{eq:E_by_V_transform}
    \end{equation}
    where $E_i$, $V_i$ and  $E_f$, $V_f$ are the initial and final energy and volume, respectively and $f(T_i, T_f) $ is a 
    transformation function of the initial and final temperatures $T_i$ and $T_f$, respectively.  For a black body
    of photons the total energy  is given by  $E = u(T) V$ where $u(T)$ is the black body energy density (\ref{eq:u(T)}) 
    at temperature $T$ and V is its volume.  Applying this to our model (\ref{eq:E_by_V_transform}) and after
    manipulation we have
    \begin{equation}
       f(T_i, T_f) = \frac{T^4_i} {T^4_f}. \label{eq:temperature-transform}
    \end{equation}
    This is the transformation law that preserves the black body distribution as the radiation expands\cite{tolman-1}.
    Since we are interested in the volume change,  substitute (\ref{eq:temperature-transform})
    into (\ref{eq:E_by_V_transform}) and manipulate it to get the result
    \begin{equation}
      V_f = \left( \frac{T^4_i} {T^4_f}  \right) \frac{ E_f V_i} { E_i } = \left( \frac{T^4_i} {T^4_f} \right) \frac{E_f} { u(T_i) },
             \label{eq:V_f_from_V_i}
    \end{equation}
    where we made the substitution $u(T_i) = E_i / V_i$.
    Since $E_f = (1 - g ) m c^2$, by definition of the CBR black body energy at the ionization epoch,
    with $T_i = T_o$, $T_f = T_{\gamma}$ given by (\ref{eq:T_gamma}) and $V_{\gamma} = V_f$ we obtain a volume
    \begin{equation}
         V_{\gamma}  = \left(  \frac{T^4_o} {T^4_{\gamma}} \right) \frac{\left( 1 - g \right) m c^2 } { u(T_o) },
             \label{eq:Vf_from_V_i_resolved}
     \end{equation}

      Substituting numerical values for the various parameters into (\ref{eq:Vf_from_V_i_resolved}), with $g = 0.04426$
      we obtain for the ionization volume  $V_{\gamma} \approx 4.23 \times 10^{71}  {\rm cm}^3$, which corresponds to a 
      radius of $R_{\gamma}  \approx 4.61 \times 10^{23} {\rm cm} \approx 0.149 \, {\rm Mpc}$.  The cosmological redshift
      $z$ of this ionization epoch is given by $(1 + z) = T_{\gamma} / T_o \approx 2.14 \times 10^4$.
 
\section{  Entropy of the black hole sub-universe }

   The entropy $S$ of a Bekenstein-Hawking black hole \cite{bekenstein-1,bh-entropy} is given by
    \begin{equation}
      S = \frac{k c^3 A } {4 \hbar G},  \label{eq:S_bh}
    \end{equation}
    where $k$ is Boltzmann's constant, and $A$ is the surface area of the  event horizon.  The surface area of the 
    black hole is given by $A = 4 \pi R^2$ where $R$ is the radius of the event horizon.
    The radius of the event horizon for one black hole sub-universe is given by  (\ref{eq:Rm}), where 
   $R = R_m  = c \tau$.    Then, the black hole entropy (\ref{eq:S_bh}) of
    the sub-universe is given by
    \begin{equation}
       S = \frac{ k c^3 \left( 4 \pi c^2 \tau^2 \right) } { 4 \hbar G } = \frac{\pi  k \tau^2 c^5 } { \hbar G }.
             \label{eq:S_bh_expand}
    \end{equation}
    Manipulating (\ref{eq:S_bh_expand}) we can get it into the form
    \begin{equation}
      \frac{ 1 } { \left( S / k \right) } \left( \frac{ -3 c^5  } {  8 \hbar G^2 } \right)  = \frac { -3 } { 8 \pi G \tau^2  } = \rho_{vac}. 
           \label{eq: entropic _vacuum_density }
    \end{equation} 
    The second factor on the l.h.s. of (\ref{eq: entropic _vacuum_density }) can be expressed as the cosmological 
    Planck mass density $\rho_P$ in terms of the  cosmological Planck mass ${\cal M}_P$ and  
    length ${\cal L}_P = \hbar /   {\cal M}_P  c$  defined by
     \begin{equation}
      \rho_P = \frac{-{\cal M}_P } { {\cal L}^3_ P } = \frac{ -3 c^5  } {  8 \hbar G^2 }.  \label{eq:Planck_vac_dens}
     \end{equation}  
     Note that for clarity we have attached the negative sign outside of the mass symbol but the implication is for  a negative mass ${\cal M}_P$. 
     Solving (\ref{eq:Planck_vac_dens}) for ${\cal M}_P$ gives
     \begin{equation}
       {\cal M}_P = \sqrt{ \sqrt{\frac{3}{8}} \frac{\hbar c } {G}}.    \label{eq:Planck_mass}
     \end{equation}
     Substituting $\rho_P$  into (\ref{eq: entropic _vacuum_density }) and rearranging we have for the vacuum mass density
    \begin{equation}
      \rho_{vac} =  \frac{\rho_P } { \left( S / k \right) } .    \label{eq:rho_vac_rho_P}
    \end{equation}
    Substituting values for the parameters in (\ref{eq:S_bh_expand}) we obtain
    \begin{equation}
      \left( S / k \right) \approx 1.980 \times 10^{122}.     \label{ewq:Sbh_by_k}
    \end{equation}
   The standard Planck mass is $m_{P} = \sqrt{\hbar c / G}$,  so that the cosmological Planck mass 
    $ {\cal M}_P  \approx 0.7825 \, m_{P}$.

\section{Discussion  \label{discussion}}

   Since no physical laws are violated during the particle, antiparticle production process
   which we  have  modeled, the particles are deemed to be real rather than virtual.   Each particle with its
   event horizon constitutes a sub-universe with no mixing of matter between sub-universes, although gravitational
   interactions occur between event horizons as we have modeled.  In other words, the gravitational field of the other
   sub-universe should cause effects in our sub-universe.

   Observations of the cosmos show that our universe is composed of baryons and leptons
   with undetectable amounts of antibaryons and antileptons.  In the standard big bang theory this indicates a
   violation of the conservation of baryon and lepton numbers.  Standard physics does not have a clear cut reason 
   for this violation \cite{cohen-1}.  In our model, both particles in separate sub-universes take part in ensuring that
    the quantum numbers are conserved across universes. The particle in our sub-universe would have been  composed
   of a mixture of more quarks, $(x+y) q$, than antiquarks, $x {\bar q}$,
   and visa-versa for the black hole particle in the other sub-universe.

\section{Conclusion}

  We assumed that the initial state of the universe was empty,  having zero energy.  We hypothesized that a 
  particle and its antiparticle were produced along with the vacuum.  Similar to  the standard model, matter has
  positive energy,  but energy conservation requires in our model that the vacuum have negative
  energy.  In the weak gravity field model, by conservation of energy and Newton's laws we were able to
  derive the mass of each  particle $m = \tau  c^3 / 4  G$, the initial separation between the particles
   $r_0 = c  \tau / 2$  and  the vacuum mass density  $\rho_{vac} = -\rho_c = -3/8 \pi G \tau^2$.
   Although there is no cosmological constant in CGR, from the vacuum density we get a corresponding expression 
   for the standard model cosmological constant $\Lambda = 3 / \tau^2$.

   By analyzing the CMB radiation energy and comparing it to the rest energy of the original black hole mass
   we found a relation between the radiation and  particle rest energy.
   Since the average radiation energy $\epsilon_{\gamma} \approx 13.6 \,{\rm eV}$, which
   is the ionization energy of the Hydrogen atom,
   we made the assumption that $\epsilon_{\gamma}$ is exactly equal to the first quantum level ionization energy
   of the Hydrogen atom when defined as
   $\epsilon_{\gamma} = ( 1 - g ) m \, c^2 / N_{\gamma} = \alpha^2 \,\mu \,c^2 / 2$, with $g$ related to
   the baryon density parameter $\Omega_b$.  We were able to express the parameters $\tau$ and $\Lambda$
   in terms of $\Omega_b$, but this does not imply that they are dependent on that parameter.  Instead, the 
   dependent parameter is the CMB temperature $T_o$,  for which by rearranging  (\ref{eq:tau_formula}) we have
    \begin{equation}
      T_o = \left[ \frac{ 45 \zeta \left( 1 - g \right) c^3 \hbar^3} {4 \pi^3 G \mu \alpha^2 \tau^2 k^3}    \right]^{1/3}. 
         \label{eq:To_formula}
    \end{equation}
   With the published value of $\tau = 4.28 \times 10^{17} {\rm s}$ and for $g = 0.04426$ 
   we get a predicted value of $T_o = 2.67 \, {\rm K}$, which is $98 \, \%$ of the experimental value.

  Finally, by investigating black hole thermodynamics we derived the Bekenstein-Hawking entropy $S$ of the black hole
  sub-universe and from that  we determined the relation  between the vacuum mass 
  density and the cosmological Planck vacuum density, $\rho_{vac} = \rho_P / S / k$ .
  Is this relevant to the cosmological constant problem of the standard model?

\section*{Acknowledgments}
   In gratitude and memory of Moshe Carmeli who gave us his cosmology.
   And thanks to my college mentors Victor J. Stenger and Sandip Pakvasa, particle physicists.

\end{document}